  \documentclass [12pt,a4paper      ]{article}
\usepackage{graphics}
\usepackage{times}

\DeclareFontFamily{OT1}{times}{}
\DeclareFontShape {OT1}{times}{m }{n }{ <-> ptmr }{}
\DeclareFontShape {OT1}{times}{bx}{n }{ <-> ptmb }{}
\DeclareFontShape {OT1}{times}{m }{it}{ <-> ptmri}{}
\DeclareFontShape {OT1}{times}{bx}{it}{ <-> ptmbi}{}
\usepackage{amsmath}
\usepackage{amsfonts}
\usepackage{amssymb}
\usepackage{latexsym}
\begin{document}

\title{\bf\vspace{-2.5cm}                The B61-based\\
                                ``Robust Nuclear Earth Penetrator:''\\
                             \emph{Clever retrofit or headway towards
                               fourth-generation nuclear weapons?} \footnote{
                             Funding for a nuclear ``bunker-buster'' warhead
                             has been dropped from the U.S.\ Energy Department's
                             2006 budget, according to information released
                             on October 25, 2005.  However, like other major
                             political issues, the project of earth penetrating
                             warheads is a recurring subject, which has
                             to be assessed in the context of an evolving
                             technological and strategic environment, that
                             is becoming particularly complex because of the
                             advent of fourth-generation nuclear weapons and
                             new nuclear-weapon States.} }

\author{{\bf Andre Gsponer}\\
\emph{Independent Scientific Research Institute}\\ 
\emph{ Box 30, CH-1211 Geneva-12, Switzerland}\\
e-mail: isri@vtx.ch}

\date{Version ISRI-03-08.18 ~~ \today}

\maketitle

{\begin{abstract}

It is scientifically and technically possible to build an earth penetrating device that could bury a B61-7 warhead 30 meters into concrete, or 150 meters into earth, before detonating it.  The device (based on knowledge and technology that are available since 50 years) would however be large and cumbersome.  Better penetrator materials, components able to withstand larger stresses, higher impact velocities, and/or high-explosive driven penetration aids, can only marginally improve the device.  It is concluded that the robust nuclear earth penetrator (RNEP) program may be as much motivated by the development of new technology directly applicable to next generation nuclear weapons, and by the political necessity to periodically reassess the role and utility of nuclear weapons, than by the perceived military need for a weapon able to destroy deeply buried targets. 

~~\\
~~\\
~~\\

\end{abstract}}

\section{Introduction}
\label{int:0}
 	
Nuclear weapons are the most destructive instruments of war available to the military.  Their destructiveness is due to a number of effects --- mechanical, thermal, and radiological --- but by far the most important military effect of present-day nuclear weapons is due to the enormous shock-waves generated in the surrounding materials by the energy released from nuclear fission and thermonuclear fusion within the warhead.

In an atmospheric explosion the shock-waves generated in air propagate over large distances, destroying every building and laying down every tree within about 3 km  from the point of explosion of a 1 \emph{Megaton}\footnote{We use italics when \emph{tons}, \emph{kt}, or \emph{Mt} refer to tons of TNT equivalent.} thermonuclear weapon.  It is this air-blast shock-loading effect combined with the incendiary effect of high-yield thermonuclear explosion which is at the origin of the strategy of war deterrence by mutually assured destruction.

However, if instead of striking relatively weak structures such as buildings the air-blast strikes the ground, most of the energy in the shock-waves is reflected so that very little energy is in fact transferred to the ground.  This very poor ``coupling'' of the energy released by an atmospheric explosion is due to the factor of about 1,000 difference in densities between the air and the soil, so that because of energy and momentum conservation the maximum energy coupling to the ground is at most of 15\% \cite[p.195]{BRODE1968-}. More precisely, in the case of low-altitude near-surface explosions (which maximize the energy transfer to the ground) the typical theoretical energy couplings are between 5 to 10\% for chemical explosions, and about 8\% for nuclear explosions --- provided the direct  coupling of X-rays  (about 6\%) is added to the mechanical coupling (about 2\%) \cite[p.874]{KNOWL1977-}.\footnote{In principle one has to include the energy of the neutrons escaping from the warhead.  This contribution to the coupling is negligible for first and second generation nuclear weapons.  However, for pure-fusion third and fourth generation nuclear weapons, energy transfer by neutrons becomes predominant.  For such weapons the coupling to the ground or a target can therefore be large, even if the weapons do not penetrate into them \cite{GSPON2005-}.}

Consequently, if the objective of a weapon is to destroy underground targets by means of shock-waves, the obvious idea is to detonated it under the surface of the ground.  This enables a reduction by a factor of about 10 to 15 in the explosive yield of the weapon.  Indeed, even for a rather shallow burial (about 0.6 meter), the maximum energy coupled to the ground by a 0.5~\emph{kt}  fission-explosion was computed to be already about 30\% of the total yield, an estimate that was confirmed by a nuclear test conducted at the Nevada Test Site \cite[p.899]{ORPHA1977-}.  Therefore, nearly 100\% coupling is achieved when a warhead is detonated at a few meters below the ground, so that the explosion of an earth penetrating warhead (EPW) with a yield of about 0.5~\emph{Mt} can in principle have the same destructive power as a 5~\emph{Mt} above ground detonation.

The military rationale for developing and producing the B61-11 (that is the 11th modified version of the B61 gravity bomb whose development started in 1961) was therefore to provide a replacement for the 9~\emph{Megatons} B53 that was intended to be used for destroying buried command centers and other underground targets.  Moreover, since the precision and the flexibility of the delivery of the B61 were substantially greater than those of the B53, it was felt that these advantages would more than compensate for the lower yield of the B61, which is at most of 340 \emph{kt}.

\begin{figure}
\begin{center}
\resizebox{8cm}{!}{ \includegraphics{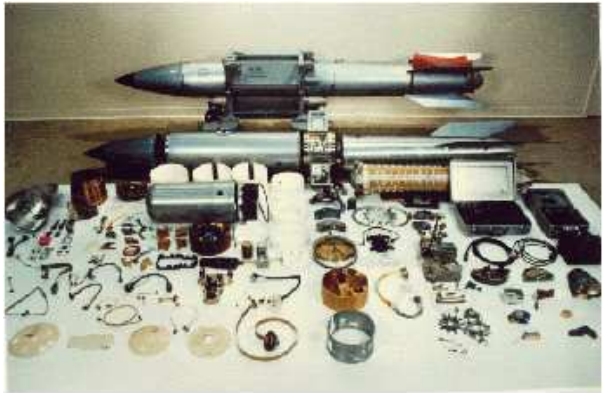}}
\end{center}
\caption{\emph{A number of components and a partially disassembled B61 bomb are shown in front of a fully assembled weapon (rear). The cylinder with a rounded head on the left, located in the center subassembly of the B61, is the ``physics package'' containing the actual thermonuclear warhead.}}
\label{fig:0}
\end{figure}

The B61-11 began entering service in 1997. It was therefore the first ``new'' nuclear weapon to enter into the arsenal of a declared nuclear power after the conclusion (in 1996) of the negotiations of the Comprehensive Test Ban Treaty, during which the United States stated that the B61-11 would be the last weapon to enter into its nuclear arsenal.  So why is it that the United States's administration is now seeking a lifting of the 1993 ban on development of new nuclear weapons (the Spratt-Furse Law)?  On 9 May 2003 a committee of the United States Senate passed a bill to that effect, calling for the development of a  ``bunker buster'' bomb to be called the robust nuclear earth penetrator (RNEP).

Part of the reasons are of a military-technical nature: Even though the B61-11 can bury itself 3--6 meters underground before detonation, it cannot destroy targets at depths or distances much larger than 30--100 meters below or away from the point of explosion.  This is because shock-waves propagating through a solid medium are much less destructive that those propagating through air \cite[Chap.6]{GLASS1977-}.  For instance, ordinary buildings and even much stronger structures such as bridges or above-ground military constructions are relatively brittle: The interaction of the blast-wave from an atmospheric explosion with them will be essentially inelastic, resulting in their complete failure at relatively large distances.

On the other hand, the interaction of an underground shock wave with a tunnel or a buried bunker will be elastic over a considerable range of overpressure, with the result that well-designed man-made structures will not fail unless the materials themselves fail.  For example, the assured destruction range of a shallow-buried explosion of 1~\emph{Mt} is on the order 150--250~m.  This corresponds to the onset of gross rock failures for most formations and thus represents the range of survival for the best examples of underground construction \cite[p.200]{BRODE1968-}.  Assuming a scaling with the third root of the yield, this confirms the stated maximum destruction range of the B61-11, i.e., about 100 meters.  However, this maximum destruction range does not take into account the protection measures that can be taken in order to improve the resistance of underground structures to the adverse effect of shock-waves.  These include many techniques such as decoupling by means of spaced or multilayered walls made of materials of different densities; shock absorption, reflection or diffusion; shock mounts for fragile equipments; etc.  As a result, one can easily estimate that the stated lower destruction range of 30~m is truly a minimum, even for a 340~\emph{kt} explosion, because some carefully designed structures may survive even that close to the point of the explosion.\footnote{Assuming a third root scaling, a factor of 3 in distance corresponds to a damping by a factor of about 30, which is not very difficult to achieve with proper civil engineering.}

It is therefore possible, on the basis of these purely technical facts, and without addressing the related military and political implications, to argue that a warhead that would be able to bury deeper into the ground before exploding would be more effective to destroy underground targets.  However, even with perfect accuracy in weapon delivery, this ability to explode at a greater depth will only get the point of explosion closer to the target --- It will not be able to extend the lethality range of the weapon itself.  For example, if the minimum assured lethality range of the B61-11 is 30--100 m, the corresponding range for a RNEP version able to bury 30~m deeper into the ground \cite{HAMBL2002-, GUINN2003-, MALAK2003-, HAMBL2005-} will be 60--130 m. Therefore, contrary to certain claims, it seems impossible that a RNEP based on the B61-11 could crush targets 300~m underground.  This is because extending the penetration range of the B61-11 by a factor of 5--10, that is from 3--6 m to about 15--30 m in reinforced concrete or solid rock, is already a considerable technical challenge, as will be seen.\footnote{In this paper we consider only the mechanical effects of nuclear weapons on buried targets.  The use of penetrating nuclear weapons against targets containing chemical or biological agents has been the subject of a number of reports, which are generally skeptical on the effectiveness of this type of engagement.  See, e.g., \cite{MAY---2004-}.}

   The paper is organized as follows:

In the Sec.~\ref{fun:0} we review the physics of the penetration of a strong and rigid projectile into a compressible solid target in order to find out whether there is room for improvement in comparison to what is known to be possible with existing materials and techniques. After deriving from first principles the rather simple expression giving the maximum hypothetical penetration depth as a function of the key physical parameters, Eq.~\eqref{fun:12}, it is found by comparisons with a dozen different penetration experiments (in which a variety of projectile and target materials are used) that there is very little room for improvement: not more than 30 to 60\%.

Starting form Sec.~\ref{pen:0} we specialize to a concrete target, for which Eq.~\eqref{pen:1} gives a good estimate of the penetration depth as a function of the length, effective density, and velocity of the penetrator.  It is found that the stated penetration capability of the B61-11 is consistent with its published physical mensurations and weights.

In Sec.~\ref{rob:0} we give the physical and engineering reasons explaining the B61-7 impact resistance (the B61-7 is the warhead of which the B61-11 is a ruggedized earth-penetrating version), which enables it to withstand the considerable deceleration upon impact with the ground.  We also derive simple formulas giving estimates for the maximum and mean decelerations during penetration.

In Sec.~\ref{sci:0} we assess the scientific feasibility of a RNEP able to bury the B61-7 warhead 30 meters into concrete, or 100 to 150 meters into earth.  Four general possibilities are analyzed: \emph{(i)} maximizing the weight and length of the penetrator; \emph{(ii)} maximizing the velocity of the penetrator; \emph{(iii)} using conventional penetration aids such as a precursor shaped-charge jet; and \emph{(iv)}, assuming that radically new types of compact nuclear explosives will become available in the next few decades, speculations are made on hypothetical new designs which could provide a technically more attractive solution than the three previous possibilities. 

In Sec.~\ref{env:0} we briefly recall the environmental and political limitations to the use of nuclear weapons, even when detonated underground.  In particular, we stress the political and nuclear proliferation problems posed by the ``plutonium mines'' created when an underground plutonium bomb explosion is contained.

In the last section, Sec.~\ref{dis:0}, we conclude that the technology for building a working but cumbersome penetrator suitable to bury the B61-7 warhead 30 meters into concrete is available, but that improving that penetrator and trying to reduce its size and/or weight by a factor of two will be very difficult, even if considerable advances are made in the realms of materials science,  micro-electromechanical systems's engineering, and nanotechnology.  Finally, considering these difficulties, as well as the military and political limitations to the battle-field use of existing types of nuclear weapons, we wonder whether apart from keeping the nuclear weapons debate alive, an important reason for considering a new generation of nuclear earth penetrating weapons would not be to develop their associated non-nuclear technologies, which overlap with those required for radically new types of weapons, both conventional and nuclear \cite{GSPON2005-}.

\section{Fundamental physical limits to kinetic penetration}
\label{fun:0}

A recurring suggestion in the general scientific literature is that by some new ``trick'' the penetration depth of a kinetic penetrator could be substantially increased (perhaps by a factor of ten or more) over what is presently possible with available materials and technologies, e.g., \cite{HAMBL2005-}.  In order to show that this is not possible, we will study a simple model in which everything is idealized except for the most basic processes which enable a high-velocity object to penetrate into a material such as steel, concrete, or earth.  Moreover, the model will be such that all approximations tend to overestimate the penetration depth, what is for instance the case for a one-dimensional model in comparison to a two- or three-dimensional one. 

For this model we follow the general idea suggested in Ref.~\cite{LAVRE1980-}, namely that when a high-velocity projectile penetrates into a material it creates a shock wave which compresses and heats the material ahead of it: The maximum penetration depth is then obtained when the material immediately ahead of the projectile (i.e., sufficiently heated to evaporate) is instantaneously removed by some ideal process, so that the penetrator can continue its course until its kinetic energy is exhausted, or some fundamental limit is reached.  In other words, the basic idea is that the kinetic energy of the penetrator is entirely used to melt the material ahead of it, and that all other processes which remove that material are supposed to have no energy cost.  Similarly, it is supposed that the frictional forces on all sides of the penetrator are negligible, and that all flows of materials in the regions adjacent to the surface of the penetrator are completely free.  However, whereas the velocity is supposed to be high, it is not supposed to be so high that one is in the high-velocity hydrodynamic-limit where all materials (including the penetrator which in the present model is supposed to be perfectly strong and rigid) behave as ideal incompressible-liquids, and all the kinetic energy is used to move them aside from the path of the penetrator.  

Let us therefore take a penetrator of length $L$, frontal area $S$, density $\rho_p$, and very high compressive strength $Y_p$; and a semi-infinite target of density $\rho_t$ and compressive strength $Y_t$.\footnote{To be precise,  what matters most in the present problem is not the ``bulk compression modulus'' $K$, but the ``unconstrained (or unconfined) compressive strength.'' For metallic projectiles and cohesive targets it has a magnitude comparable to the yield strength $Y_s$, which is typically in the range of 10~Mpa to 2~GPa.  For cohesionless or granular targets such as soil or sand, the unconstrained compressive strength in on the order of 0.1 to 10~MPa.}  To begin with we suppose that this penetrator is incompressible and sufficiently strong to remains intact during penetration into the target.\footnote{That restriction will be removed at the end of this sections, when the maximum impact velocity consistent with our model will be evaluated.} According to our model, at any given time during penetration, the velocity of the projectile $v_p(t)$ is such that the compressive strength of the target is exceeded, so that ahead of it the target material is compressed by a shock-wave to a density $\rho_s > \rho_t$. As we are considering large penetration velocities, we can assume the strong shock limit, in which case 
\begin{equation}\label{fun:1}
                            \rho_s = \rho_t \frac{1+\gamma}{1-\gamma},
\end{equation}
where $\gamma > 1$ is the polytrope exponent.

The crucial step in the reasoning is now to apply energy-momentum conservation to the shocked target material ahead of the penetrator, which implies
\begin{equation}\label{fun:2}
                  d(\rho_s S x v_s)  = \rho_s S ( x \, dv_s + v_s \, dx ) = 0,
\end{equation}
where $x$ is the distance measured (for convenience) from the back of the penetrator and $v_s(x,t)$ the velocity of the shocked material at any given time.  With the boundary condition $v_s(L,t)=v_p(t)$, this equation is trivially integrated, i.e.,
\begin{equation}\label{fun:3}
                 v_s(x,t) = v_p \frac{L}{x},
\end{equation}
One can therefore calculate the energy distribution in the shocked material
\begin{equation}\label{fun:4}
            E_s(x,t) = \int_L^x \frac{1}{2} S \rho_s v_s^2 ~dx
                   = E_p \frac{\rho_s}{\rho_p} (1 - \frac{L}{x}),
\end{equation}
where $E_p(t) = \frac{1}{2} S \rho_p v_p^2(t)$ is the kinetic energy of the penetrator at time $t$.  By derivating this equation and dividing by $S$, or by directly using \eqref{fun:3}, one can also calculate the energy density, i.e.,
\begin{equation}\label{fun:5}
            \frac{d\,E_s}{dV}(x,t) =  \frac{1}{2} \rho_s v_s^2
                   = \frac{1}{2} \rho_s v_p^2 \frac{L^2}{x^2}.
\end{equation}

The last equation can now be used to write down the \emph{criterion for penetration}, which is simply the condition that the energy density in front of the penetrator has to be higher than that corresponding to the compressive strength of the target, i.e.,
\begin{equation}\label{fun:6}
            \frac{d\,E_s}{dV}(x,t)
                   = \frac{1}{2} \rho_s v_p^2 \frac{L^2}{x^2} \geq Y_t.
\end{equation}
In particular, the \emph{minimum velocity}\footnote{This velocity is not the same as the usual estimate for penetration to occur, which is derived from the ratio of the projectile ram pressure $\tfrac{1}{2}\rho_p v_p^2$ to the target compressive strength, i.e., Eq.~\eqref{fun:7} with $\rho_s$ replaced by $\rho_p$.} of the projectile for penetration to proceed until it has come to a rest is given when $x=L$, i.e.,
\begin{equation}\label{fun:7}
            v_{p,\text{min}} = \sqrt{\frac{2Y_t}{\rho_s}}.
\end{equation}

Equation \eqref{fun:6} can also be used to calculate the thickness $\Delta z = x - L$ of the shocked region such that its energy density exceeds $Y_t$, i.e., 
\begin{equation}\label{fun:8}
            \Delta z + L = L \sqrt{\frac{\rho_s}{2Y_t}}.
\end{equation}
In the present model, the energy $\Delta E = E_s(\Delta z + L,t)$ given by Eq.~\eqref{fun:4} corresponds to the energy used to ``evaporate'' the material in the region of thickness $\Delta z$ ahead of the penetrator.  Therefore, since that material is supposed to be instantaneously removed out of its way, the penetrator's linear energy loss is given in first approximation by
\begin{equation}\label{fun:9}
         \frac{dE_p}{dz} \approx -\frac{\Delta E}{\Delta z}
        = -\frac{E_p}{L} \frac{\rho_s}{\rho_p}
        \sqrt{\frac{2Y_t}{\rho_s}} \frac{1}{v_p},
\end{equation}
or, expressing $v_p$ in terms of $E_p$ and rearranging, by
\begin{equation}\label{fun:10}
   \frac{dE_p}{2\sqrt{E_p}} = - \sqrt{\frac{S}{L} \frac{\rho_t}{\rho_p} Y_t} ~ dz,
\end{equation}
where, because the removed material is in gaseous form, we have approximate its adiabatic exponent by that of a monoatomic gas, i.e., $\gamma = 5/3$, so that Eq.~\eqref{fun:1} gives $\rho_s=4\rho_t$.

Equation \eqref{fun:10} is easily integrated by quadrature, giving the penetration depth $D=z$ in function of the initial and final penetrator energies
\begin{equation}\label{fun:11}
      D = \sqrt{ \frac{L}{S} \frac{\rho_p}{\rho_t} \frac{1}{Y_t}}
                   \Bigl( \sqrt{E_p}(0) -  \sqrt{E_p}(1) \Bigr),
\end{equation}
or in function of the initial and final penetrator velocities
\begin{equation}\label{fun:12}
      D = L \frac{\rho_p}{\rho_t} \sqrt{ \frac{\rho_t}{2Y_t} }
                 \Bigl( v_p(0) -  v_p(1)  \Bigr),
\end{equation}
where the final velocity $v_p(1)$ can been replaced by the velocity given by Eq.~\eqref{fun:7}, i.e., assuming again $\gamma = 5/3$,
\begin{equation}\label{fun:13}
            v_p(1)= \sqrt{ \frac{Y_t}{2\rho_t} }.
\end{equation}

  The main result of our model, equation~\eqref{fun:12},\footnote{While dimensional analysis directly leads to its general form, and many papers use it in a form or another, I have not found a published derivation from first principles of this equation.} is remarkable because it explains essentially all the important qualitative features observed in penetration experiments: direct proportionality to the length, density, and velocity of the penetrator; inverse proportionality to the square root of the density and to that of the compressive strength of the target; and existence of a minimum velocity.

   On the other hand, when making quantitative comparisons with the data, it turns out --- as expected form the discussion in the introduction to this section --- that Eq.~\eqref{fun:12} tends to overestimate the penetration depth.  This can be seen, for example, if one looks at the penetration depth versus striking velocity of steel projectiles into concrete data of Ref.~\cite{ROGER1996-} shown in Fig.~2 of Ref.~\cite{NELSO2002-}.  In that case Eq.~\eqref{fun:12} overestimates $D$ by a factor of $\approx$ 2.8.  But, if one makes the comparison with the data of Ref.~\cite{CANFI1966-} shown in Fig.~7 of Ref.~\cite{BERNA1976-}, the discrepancy is only a factor of $\approx$ 1.5 (see, Table~\ref{tbl:1}).  Consequently, although the details of the experimental conditions and the exact properties of the materials used are not well known, the ``hypothetical maximum room for improvement'' suggested by these comparisons is small: at most between a factor 1.5 and 3.  To better understand what this means in practice, further comparisons with other experiments using other target and projectile materials are made in Table~\ref{tbl:1}.
\begin{center}
\begin{table}
\begin{tabular}{|c|c|c|c|c|c|c|c|c|}
\hline
\multicolumn{9}{|c|}{\raisebox{+0.4em}{\bf Comparison of ideal maximum penetration formula, Eq.~\eqref{fun:12}, to data \rule{0mm}{6mm}}} \\
\hline   
{\bf ~}  &  & $\rho_t$  & $Y_t$ & $D/L$ & $d/l$ & $v_{\text{f}}$ & $D/L$ &  R   \rule{0mm}{5mm}\\
~        &  & [kg/m$^3$ & [MPa] & (exp) & [s/km]&   [km/s]  & \eqref{fun:12} &           \\
\hline
Concrete &\cite{CANFI1966-}&   2310    &  34.7 &   12  &  20   &    0.09   &    18 &  1.5 \rule{0mm}{5mm}\\
\hline
Concrete &\cite{FORRE1996-}&   2300    &  62.8 &  6.1  &  15   &    0.11   &    13 &  2.1 \rule{0mm}{5mm}\\
\hline
Concrete &\cite{FORRE1994-}&   2340    &  96.7 &  3.7  &  12   &    0.14   &    10 &  2.7 \rule{0mm}{5mm}\\
\hline
Concrete &\cite{ROGER1996-}&   2500    &   50  &    5  &  16   &    0.10   &    14 &  2.8 \rule{0mm}{5mm}\\
\hline
Steel    &\cite{CONTI1977-}&   7900    & 1000  &    1  &  3.8  &    0.25   &   2.8 &  2.8 \rule{0mm}{5mm}\\
\hline
Titanium &\cite{CONTI1977-}&   4500    &  800  &  1.2  &  5.5  &    0.30   &   3.9 &  3.2 \rule{0mm}{5mm}\\
\hline
Lexan    &\cite{CONTI1977-}&   1210    &   60  &   10  &  40   &    0.16   &    34 &  3.4 \rule{0mm}{5mm}\\
\hline
Aluminum &\cite{CONTI1977-}&   2700    &  100  &   4   &  21   &    0.14   &    18 &  4.5 \rule{0mm}{5mm}\\
\hline
Grout    &\cite{FORRE1996-}&   2000    &  21.6 &  7.6  &  27   &    0.07   &    25 &  3.3 \rule{0mm}{5mm}\\
\hline
Grout    &\cite{FORRE1996-}&   2000    &  13.5 &  8.1  &  34   &    0.06   &    32 &  4.0 \rule{0mm}{5mm}\\
\hline
Lead     &\cite{LAVRE1980-}&  11300    &   17  &   2.5 &   3   &    0.03   &    13 &  5.2 \rule{0mm}{5mm}\\
\hline
Foam     &\cite{LAVRE1980-}&    110    &   20  &  120  &  120  &    0.30   &    85 &  0.7 \rule{0mm}{5mm}\\
\hline
\end{tabular}
\caption{\emph{Ratio $R=(D/L)_{\text{hyp}} /(D/L)_{\text{exp}}$ normalized to an impact velocity of $v_p=1$~km/s, where  $(D/L)_{\text{hyp}} = d/l(v_p - v_f)$ is calculated with Eq.~\eqref{fun:12}, and $(D/L)_{\text{exp}}$ is measured for the materials listed in column 1 to 4. The symbol $d/l$ is the slope coefficient, and $v_f$ the final penetrator velocity given by Eq.~\eqref{fun:13}.}}
\label{tbl:1}
\end{table}
\end{center}

\vspace{-1.4cm}

   Let us define the hypothetical room for improvement by the ratio $R$ between the hypothetical maximum penetration given by Eq.~\eqref{fun:12} and the corresponding measurement for a given penetrator length $L$ at a normalized penetrator velocity of $v_p(0) = 1$~km/s, i.e., $R=(D/L)_{\text{hyp}} /(D/L)_{\text{exp}}$ at that velocity.  For the sake of comparison, and in order to exhibit some possible trends, the data in Table~\ref{tbl:1} is listed in approximately increasing order of $R$, except for the last experiment in which the target is a low density foam.  

   The first set of data, from Refs.~\cite{CANFI1966-, FORRE1996-, FORRE1994-, ROGER1996-}, is for steel (i.e., $\rho_p = 7.9$~Mg/m$^3$) ordnance-type projectiles into concrete targets.  As can be seen, this set corresponds to the smallest room for improvement, possibly because everything feasible has already been done with them in order to facilitate penetration, e.g., maximal hardness, ogival nose shape, and large length over diameter ratio (``long rod'' penetrators).

   The second set of data, from Ref.~\cite{CONTI1977-}, is for tungsten carbide (i.e., $\rho_p = 15$~Mg/m$^3$) spherical balls into various typical armor materials: rolled homogeneous armor (RHA) steel, titanium, polycarbonate (G.E.\ Lexan), and 1100-F aluminum.  As can be seen $R$ is comprised between 2.8 and 4.5, with a slight trend towards an increase when going from harder to softer materials.  Compared to the previous set of data, $R$ is about 1.6~times larger on average.   However, if instead of being spherical the penetrator's shape would have been optimized, and thus its drag coefficient minimized, $R$ could have been significantly smaller.  This effect is well known and accounts for a factor of about 2 when going from a flat to a conical shape (see, e.g., \cite[p.6]{NELSO2002-}).\footnote{This effect remains significative in the 1~to 2~km/s velocity range where going from a flat to a sharp nose accounts for a factor 1.4 in penetration depth \cite{ROSEN1999-}.}

   The third set of data consists of experiments with ogive-nose steel projectiles impacting grout with two different compressive strengths \cite{FORRE1996-}.  The trend for an increase of $R$ when going from harder to softer materials is observed again.

   The last set of data, from Fig.~116 of Ref.~\cite{LAVRE1980-}, is for spherical balls of steel into a soft and heavy material (lead), or a low density foam (assumed to be polyethylene).  In the case of lead the hypothetical room for improvement is 5.2, suggesting that for this material the idealizations leading to Eq.~\eqref{fun:12} are not adequate: in particular, the compressive strength is so small that little energy is needed to crush it, while significant energy is needed to sweep it aside.  Finally, in the case of the low-density foam, the material can no more be considered as a ``solid,'' even at the relatively high impact velocities considered here: Eq.~\eqref{fun:12} fails and another model should be developed.\footnote{Although $R \approx 0.7$ is not a bad theoretical estimate, the large value of $v_{\text{min}}$, especially in comparison to that of lead, suggests that a different model is definitely required.  In effect, $v_{\text{min}}$ is the minimum velocity for compressive-yielding to be the dominant penetration mechanism, while for a soft material penetration occurs at a much lower velocity.} 

  Consequently, by taking into consideration the most obvious detrimental effects that were neglected in the derivation of Eq.~\eqref{fun:12} (such as the energy needed to moved the target material out of the way, and the influence of the penetrators's nose shape) the data in Table~\ref{tbl:1} show that instead of a room for improvement of about 3 to 4 as suggested by a naive interpretation, the actual margin cannot be much more than a factor of about 1.5, or possibly even less considering the complexity and number of small effects that should be considered at the detailed level.

   In particular, this implies that processes such as cavitation, which can have a significant impact in aerodynamics and low-velocity hydrodynamics \cite{TULIN1964-}, will not be able to play any significant role in improving penetration into a solid target \cite{HAMBL2005-}, except possibly when driving into soil or mud.  Similarly,  using for the penetrator a composite or nano-engineered material instead of steel \cite{MURPH2003-}, or a heavy-material like uranium (which by adiabatic shearing enables to improve long-rod penetration into a metal\footnote{But not into concrete or rock.} by about 10\% at high-velocities \cite{ROSEN1999-,GSPON2003-}), will only have a marginal effect on penetration depth apart from that due to their densities.

   Before summarizing, let us recall that the purpose of our model was not to provide an exact fit to the data, but rather to give an estimate for a hypothetical maximum penetration depth assuming that all the kinetic energy of a perfectly hard projectile is used to ``punch'' a hole through the target, i.e., that none of that energy is used to remove the material evaporated at the head of the penetrator, and that none of that energy is lost as mechanical energy or stored as elastic energy, but rather converted to heat in order to vaporize the material ahead of the penetrator.  This means that no conceivable physical process can possibly beat the maximum penetration given by Eq.~\eqref{fun:12}.  Moreover, if the projectile itself cannot be considered as perfectly hard,  Eq.~\eqref{fun:12} will provide an even larger upper limit to the actual maximum penetration depth.  This happens when the velocity of the projectile $v_p$ is so large that the ram pressure exceeds its compressive strength $Y_p$, and that it begins to erode and loose rigidity.  An estimate for that \emph{maximum velocity} is provided by going to the projectile rest frame, where the ram pressure is  $\tfrac{1}{2}\rho_t v_p^2$, giving the expression
\begin{equation}\label{fun:14}
                  v_{p,\text{max}} \approx \sqrt{\frac{2Y_p}{\rho_t}}.
\end{equation}
Then, inserting Eq.~\eqref{fun:14} in Eq.~\eqref{fun:12}, we can derive a simple formula for the maximum possible penetration at maximum projectile velocity
\begin{equation}\label{fun:15}
      D_{\text{max}} \approx L \frac{\rho_p}{\rho_t}
          \Bigl(\sqrt{ \frac{Y_p}{Y_t} } - \frac{1}{2} \Bigr),
\end{equation}
which like Eq.~\eqref{fun:12} will of course overestimate the actual maximum possible penetration depth.\footnote{For the steel into concrete data of Ref.~\cite{ROGER1996-} one finds that $(D/L)_{\text{max}} \approx 5$, while Eq.~\eqref{fun:15} gives about 11, i.e., a factor of 2.4 more, as could be expected from the value of $R$ given in Table~\ref{tbl:1}.}

  In conclusion, even if for some improbable reason the maximum penetration depths could be larger than given by Eqs.~\eqref{fun:12} or \eqref{fun:15}  --- say by a factor of two or three --- the maximum hypothetically possible penetration depth would still not be larger than the experimentally measured and theoretically known values by one order of magnitude.\footnote{The only debatable parameter in Eq.~\eqref{fun:12} is the polytrope exponent $\gamma$ implicitly introduced through Eq.~\eqref{fun:1}.  If a reasonable value different from 5/3 is taken for this parameter, the depths given by Eqs.~\eqref{fun:12} or \eqref{fun:15} will be at most different by a factor of two.  This has, however, no influence on what is physically possible in order to improve penetration.}  Therefore, any conceivable improvement in projectile technology cannot increase penetration by a factor of be more than about 1.5.

  Finally, for the penetration depth into a solid to be maximum, the projectile velocity has to be comprised between the two values approximately given by Eqs.~\eqref{fun:7} and \eqref{fun:14}.  Below  $v_{p,\text{min}}$ penetration is not possible, and above $v_{p,\text{max}}$ penetration will decrease because of projectile erosion until reaching the hydrodynamic limit, given by the well known formula
\begin{equation}\label{fun:16}
      D_{\text{hydro}} = L \sqrt{\frac{\rho_p}{\rho_t} },
\end{equation}
which always gives a smaller penetration depth than Eq.~\eqref{fun:12}, even when modified to take compressibility effects into account, e.g., \cite{TATE-1967-, HAUGS1981-}.

\section{The penetration capability of the B61-11}
\label{pen:0}

The present earth penetrating version of the B61, i.e., the B61-11, has a length $L = 3.7$~m, and a maximum diameter $d=0.33$~m.  Its total weight is $M \approx $~550~kg, about 200~kg more than the B61-7, the model of which  the B61-11 is a modified earth-penetrating version.  Further characteristics of the B61-7 and B61-11 are listed in Table~\ref{tbl:2}, and additional information can be found in references \cite{COCHR1984-, HANSE1988-, HANSE1995-,  SUBLE1997-, NORRI2003-, NAS--2005-}.

\begin{figure}
\begin{center}
\resizebox{12cm}{!}{ \includegraphics{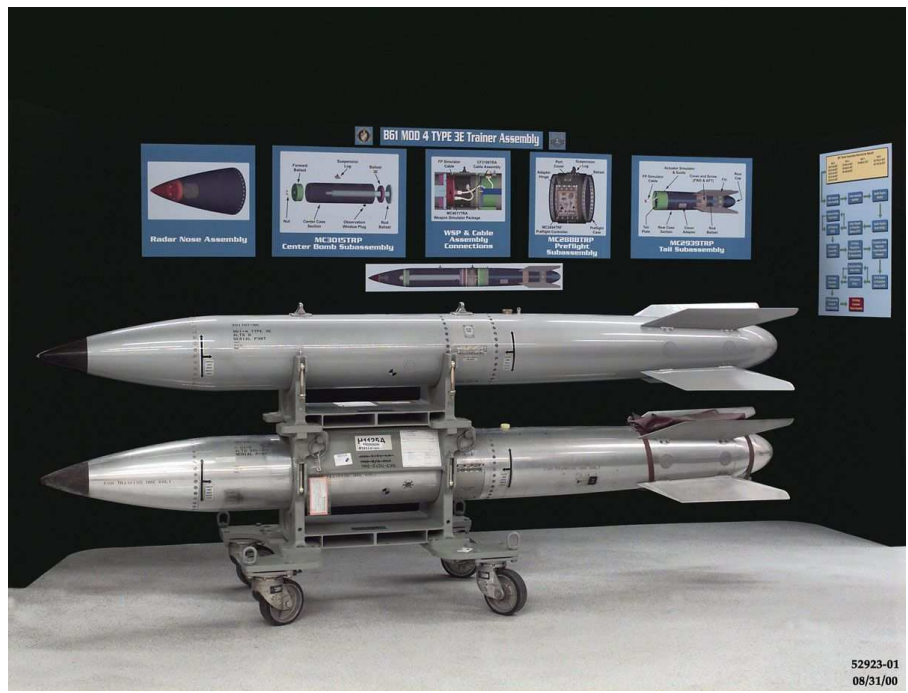}}
\end{center}
\caption{A stack of two B61-4 bombs.  The panels at the rear correspond to a training kit showing the main subassemblies of the weapon and some of their maintainable internal components.  The second panel from the left shows the ``Center Bomb Subassembly'' which houses the nuclear package.  It is shown to be traversed by a removable axial ``rod'' which corresponds to the  deuterium-tritium boost gas reservoir for the primary and the secondary stages of the thermonuclear warhead.}
\label{fig:1}
\end{figure}

   Assuming that the overall shape of the B61-11 can be approximated by an ovoid, its volume will be $V \approx (\pi/6)d^2L = 0.21$~m$^3$, which means that the effective density of the B61-11 is $\rho_{\text{eff}} = M/V \approx 2.6$~Mg/m$^3$ --- about three times less than if it was made out of solid steel.

    Let us suppose that the target material consists of concrete, for which we take the characteristics listed in the third line of Table~\ref{tbl:1}. For the sake of the argument, let us also suppose that Eq.~\eqref{fun:12}, which gives the hypothetical maximum penetration depth, is overestimating the actual penetration depth by a factor of $\approx 3$, i.e., the average value of $R$ listed in the last column of Table~\ref{tbl:1}.  Taking this factor into account, we therefore get the following expression
\begin{equation}\label{pen:1}
      D \approx  5 L \frac{\rho_{\text{eff}}}
                        {\rho_{\text{Fe} }} (v_p\text{[km/s]} - 0.1),
\end{equation}
which agrees well with the data and the two empirical formulas given in Ref.~\cite{NELSO2002-} when $\rho_{\text{eff}}=\rho_{\text{Fe}}$. 

    In Eq.~\eqref{pen:1} the penetrator's material enters by only one of its physical properties: the effective density $\rho_{\text{eff}}$.  This is an elementary but important point which, by going through the physical reasoning leading to Eq.~\eqref{fun:12}, comes from the fact that for a perfectly hard and strong projectile the only thing that matters for penetration is its total kinetic energy. 

    Another trivial but important feature of Eq.~\eqref{pen:1} is that it contains the \emph{product} of  $\rho_{\text{eff}}$ by $L$, a quantity called ``opacity'' that appears in many physical problems.  Consequently the penetration is the same if the density of a projectile is doubled at the same time as its length is halfed, provided the penetration is measured from the surface of the target to the tip of the penetrator when it comes to a rest.  This implies that is not necessary to know how the mass is distributed within the penetrator.  For example, in the case of the B61-11, one can suppose that its 550~kg are uniformly distributed over its 3.7~m of length, or else concentrated in the first half (which is actually not far from being the case, as can be seen in Table~\ref{tbl:2}) and assume that its length is only 1.85~m.

   The notion that the \emph{effective opacity} $\rho_{\text{eff}} L$ is the decisive\footnote{As we have seen in the previous section, once obvious steps such as optimizing the penetrator's shape have be taken, very little room is left for improvement.} projectile-dependent parameter defining its penetration capability implies that knowledge of the construction details of the B61-11, or any future RNEP, is not necessary.  For example, looking at the numbers in Table~\ref{tbl:2}, we see that out of the 203~kg mass difference between the B61-7 and B6-11, 133~kg are in the front.  To account for this difference several hypothesis can be made.  For example, if we approximate the nose by a cone with base diameter $d$, and assume that its made out of steel with density 7.9~Mg/m$^3$, it would weight 146~kg.  On the other hand, if we approximate the new penetration aid subassembly by a cylinder of diameter $d$, and assume that its made out of a tungsten or depleted uranium alloy of density 18~Mg/m$^3$, it would weight 153~kg. After properly adjusting the numbers, either option would lead to the same effective opacity, and thus the same penetration.  However, for many reasons, such as flight stability and terminal guidance, neither of these options will be optimum.

\begin{table}
\begin{center}
\begin{tabular}{|c|c|c|c|c|}
\hline
\multicolumn{5}{|c|}{\raisebox{+0.4em}{\bf Comparison of B61-7 and B61-11 weights and lengths \rule{0mm}{6mm}}} \\
 \hline
          
            & {\bf B61-7} & {\bf B61-11} & {\bf B61-7} & {\bf B61-11}\rule{0mm}{5mm}\\
Subassembly &     Mass    &   Mass       &   Length    &    Length    \\
            &     [kg]    &   [kg]       &    [m]      &     [m]      \\
\hline

Nose                        &            &             &   0.65  &  0.65  {\raisebox{-0.4em}{\rule{0mm}{6mm}}}\\
Center                      &            &             &   1.15  &  1.15   \\
Penetration aid             &            &             &         &  0.10   \\
Nose + center + penetrator  &     243    &      377    &   1.80  &  1.90   \\
 
\hline

Preflight controller      &     23      &     23       &   0.30  &  0.30  {\raisebox{-0.4em}{\rule{0mm}{6mm}}}\\

\hline

Tail                      &     80      &    150       &   1.50  &  1.50  {\raisebox{-0.4em}{\rule{0mm}{6mm}}}\\

\hline

{\bf Total}               & {\bf 346}   &  {\bf 549}   & {\bf 3.60} & {\bf 3.70}  {\raisebox{-0.4em}{\rule{0mm}{6mm}}}\\
\hline
\end{tabular}
\caption{\emph{The weights are from Ref.~\cite[Table~3.2]{NAS--2005-}, and the lengths are estimated from Ref.~\cite[p.166]{HANSE1988-} and  \cite{NAS--2005-}.  The maximum diameter of both bombs is $\approx 0.33$~m.}}
\label{tbl:2}
\end{center}
\end{table}

   Let us now consider the penetration capability of the B61-11. If it is dropped from a height of $h=12,500$~m, as in typical drop tests \cite{NELSO2002-}, its maximum ground impact velocity neglecting air drag would be
\begin{equation}\label{pen:2}
      v_p(0) = \sqrt{2 g h} \approx 500~\text{m/s},
\end{equation}
where $g$ is the acceleration of Earth's gravity.  

   Consequently, at a terminal velocity of 0.5~km/s, Eq.~\eqref{pen:1} gives a penetration into concrete of 2.4~m, i.e., $D/L \approx 0.65$,\footnote{This implies that the warhead should be in the front-half of the bomb, which is actually the case since it is located in the center subassembly, see Table~\ref{tbl:2}.} in agreement with the advertised capability of the B61-11.  

   How could this penetration capability be increased?  Obviously, one could increase the impact velocity by fitting the bomb with a rocket engine, and possibly reach the maximum set by the strength of the bomb casing, i.e., Eq.~\eqref{fun:15} where one has to divided by 3 to be consistent with  Eq.~\eqref{pen:1}.  For example, if the bomb's casing is made out of steel so that $Y_p \approx 1$~GPa,\footnote{In the case of maraging steel one could have $Y_p \approx 2.6$~GPa.} that maximum would be $D \approx 4.2L{\rho_{\text{eff}}}/{\rho_{\text{Fe} }}$, i.e., approximately $5$~m in concrete --- quite far away from the goal of 30~m penetration in concrete or hard rock mentioned for the RNEP in the general scientific literature \cite{GUINN2003-, MALAK2003-,HAMBL2005-}.

\section{The robustness of the B61-11}
\label{rob:0}

Before discussing a number of possible designs which could lead to greater penetration depths, it is useful to make a few comments based on elementary physical considerations, which explain how the B61-7 warhead contained in the B61-11 can withstand the enormous deceleration of impact and penetration.  This is particularly important since any possible future RNEP will most probably also use this warhead, because in the absence of nuclear tests an entirely new fission-fusion warhead cannot easily be developed, and would not have the same deterrence value as a fully tested warhead.

\begin{figure}
\begin{center}
\resizebox{8cm}{!}{ \includegraphics{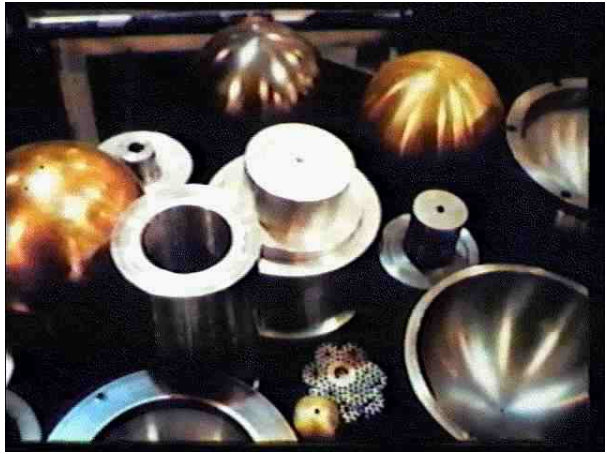}}
\end{center}
\caption{\emph{Main nuclear components of an early version of the B61 \cite{USAEC1970-}:  the fission primary (spherical and hemispherical parts), the fusion-fission secondary (cylindrical parts), and the two ends of the radiation case.  The primary has a levitated pit: the small sphere at the bottom center with a hole for the deuterium-tritium boost gas supply.}}
\label{fig:2}
\end{figure}

In order to have a quantitative characterization of the robustness of the B61-7, it is necessary to find an estimate of the magnitude of deceleration during impact.  A first approximation to this is obtained by assuming that after free-fall from a height $h$ a perfectly strong penetrator of mass $m$ and length $L$ is decelerated by a constant acceleration $a_{\text{mean}}$.  The equation $mDa_{\text{mean}} = mhg$ gives then the estimate
\begin{equation}\label{rob:1}
      a_{\text{mean}} \approx \frac{h}{D} g.
\end{equation}
For $h=12,500$~m and $D=2$~m this gives a mean deceleration $a_{\text{mean}} \approx 6,000~g$.

Is it possible that a nuclear warhead can withstand such a large mean acceleration?  The answer is certainly yes since earth penetrating warheads and nuclear artillery shells have been designed and deployed in the 1950s already.  This gives an opportunity to make an estimate for the maximum acceleration during impact, and to compare that estimate with what is known about the robustness of nuclear artillery shells.

  Suppose therefore that the ram pressure of a projectile striking a target (both assumed to be perfectly strong) is $p = \tfrac{1}{2}\rho_{\text{p}} v_p^2$, and that the pressure at impact can be written as $p = m a_{\text{max}} / S$ where $S$ is the projectile's cross-sectional area.  In the case of an artillery shell elementary ballistic physics shows that the projectile's velocity is related to the maximum range by the relation $v_p^2 = g r_{\text{max}}$.  Thus
\begin{equation}\label{rob:2}
      a_{\text{max}} \approx \frac{r_{\text{max}}}{2L} g,
\end{equation}
where $L$ is now the length of the shell assumed to be a cylinder.  If we take for example the W82 warhead, i.e., the ``neutron bomb'' artillery shell, $L= 0.87$~m and $r_{\text{max}} \approx 30$~km, giving a maximum acceleration $a_{\text{max}} \approx 17,000~g$ at impact.  In fact, by time symmetry, the argument leading to that estimate can be reversed and interpreted as the maximum acceleration during launching of the shell by an artillery gun, which in the case of the W82 is known to be 15,000~$g$ \cite{ETR--1979-}.\footnote{``The W82 warhead for the 155-mm nuclear projectile (...) must survive a 15,000~$g$ acceleration at launching'' \cite{ETR--1979-}.}  Our estimate is therefore reasonable. 

  How can a nuclear warhead survive an acceleration or deceleration on the order of 15,000~$g$?  Clearly, the warhead should contain a minimum of moving parts and all components must be mounted securely to resist shocks.  Therefore all free spaces have to be filled by a strong light-weight material such as polyurethane or beryllium, of such a density that it is compatible with the functioning of the warhead.  Moreover, since the crushing pressures are longitudinal, the warhead's design should if possible be cylindrical rather than spherical.  Finally, if the warhead is enclosed in a heavy penetrator, shock absorbers can be used to decouple it form the penetrator, and thus to reduce the stresses.

For example, the early version of the B61 shown in Fig.~\ref{fig:2} has a levitated fissile pit that must be precisely positioned at the center of the reflector and high-explosive shells surrounding it.  Because of the relatively high densities of uranium or plutonium, such a design cannot resist very large shocks, which is why most versions of the B61 are ``lay-down bombs,'' i.e., bombs that are slowed-down by a parachute and ment to explode once they are at rest on the ground.

\begin{figure}
\begin{center}
\resizebox{8cm}{!}{ \includegraphics{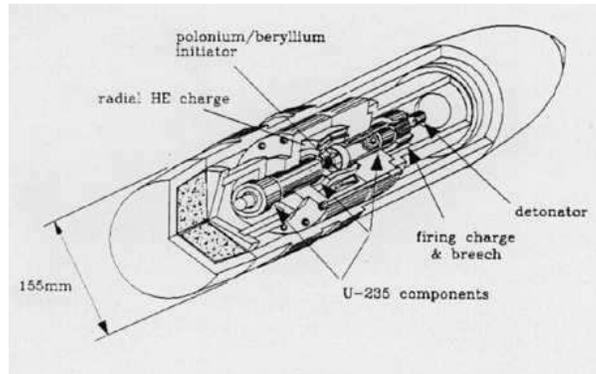}}
\end{center}
\caption{\emph{Schematic of a W48 nuclear artillery shell \cite{LARGE1995-}. A critical mass is obtained by first gun-assembling two U-235 components and then imploding them radially.  Because the overcriticality achievable is limited the yield is ``very small,'' probably 0.1~\emph{kt} \cite{COCHR1984-}.}}
\label{fig:3}
\end{figure}

\begin{figure}
\begin{center}
\resizebox{4cm}{!}{ \includegraphics{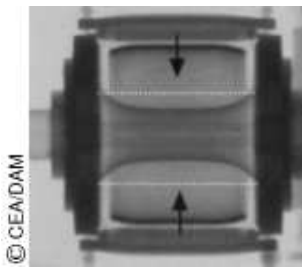}}
\end{center}
\caption{\emph{Flash X-ray radiography of the implosion of a thin cylinder  by confined high-explosives \cite{CEA--2003-}. The broken lines indicate the initial shape of the cylinder. A considerable advantage of cylindrical primaries is that a sealed deuterium-tritium boost gas reservoir can be inserted and removed axially for safety and maintenance.}}
\label{fig:4}
\end{figure}

One the other hand, the W48 artillery shell developed between 1957 and 1963 shown in Fig.~\ref{fig:3} is based on a hybrid gun-assembly/implosion design with a diameter of less than 15~cm.  This cylindrical design, code-named ``Manticore'' in Ref.~\cite{FRANC1995-}, was first tested in 1955 and opened the way to the development of cylindrical primaries for two-stage thermonuclear weapons.  It is therefore most probable that cylindrical primaries are used in the warheads of the W82 and B61-7, and therefore in the B61-11.  A French experiment related to such primaries is shown in Fig.~\ref{fig:4}.

\section{Scientific feasibility of a RNEP}
\label{sci:0}

The robust nuclear earth penetrator (RNEP) program is an engineering feasibility study under way since May 2003 in the United States, with the objective to determine if a more effective earth penetrating weapon could be designed using major components of existing nuclear weapons.  This is a quite ambitious objective because the engineering aspects of the problem are very complex.

    In this section we focus on a less difficult objective, i.e., to determine if it is scientifically possible to envisage a RNEP capable of penetrating 30~m of concrete \cite{GUINN2003-, MALAK2003-, HAMBL2005-}.\footnote{This would correspond to about 100 to 150~m of earth.}  Specifically we will consider four possibilities, and discuss them in order of increasing sophistication, which in the medium to long term may correspond to increasing order of plausibility.

\subsection{Maximization of effective opacity}
\label{sci.1:0}

According to Eq.~\eqref{pen:1}, the most simple approach to increasing penetration is to maximize the effective opacity because when dropped from a given height the terminal velocity does not depend on the penetrator's mass or size  (as recalled by Eq.~\eqref{pen:2}).  This approach was actually taken in 1991 at the end of the Gulf war to destroy a well-protected bunker north of Baghdad after repeated directed hits.  

  In that case an earth penetrating bomb, named GBU-28, was hastly created using a 3.9~m-long, 36~cm-diameter, surplus Army eight-inch-bore gun tube filled with high-explosives \cite{NELSO2001-,GRONL2002-}.  By taking densities of 7.9 and 1.9~g/cm$^3$ for the steel and the high-explosives, respectively, the weight of the GBU-28 comes out to be about 2,300~kg, and its effective density approximately 6,000~kg/m$^3$.  For a length of 3.9~m and a terminal velocity of 0.5~km/s the penetration in concrete given by Eq.~\eqref{pen:1} is 5.9~m, in good agreement with the improved version, GBU-28/BLU-113, which is claimed to ``break through 7~meters of concrete or 30~meters of earth'' \cite{HAMBL2005-}.

   Moreover, the same approach was already taken in the 1950s for designing earth-penetrating nuclear weapons \cite{NORRI2003-}.  For instance, ``the uranium gun-type Mark~8 bomb (nicknamed `Elsie' for LC, or light case) was almost 3~meter long, 36~cm in diameter, 1,500~kg, and had a yield of approximately 25~\emph{kt}. (...) The Mark~11 was an improved version of the Mark~8, slightly heavier, and according to the National Atomic Museum, `able to penetrate up to 6.6~m of reinforced concrete, 27~m of hard sand, 36~m of clay, or 13~cm of armor plate,'\footnote{There could be an error with the armor plate thickness, which should be more like 38~cm (i.e., 15 rather than 5~inches).} and fuzed to detonate 90--120 seconds after penetration'' \cite{NORRI2003-}.  In other words, the characteristics of these 1950s-bombs are very similar to those of the GBU-28, except for the nuclear warhead.

   Using the same method one can easily propose a design for a weapon able to break through 30~meters of concrete, i.e., five times more.  First, one can replace the steel tube by one made out of a strong tungsten or uranium alloy with densities of $\approx 17$~g/cm$^3$ and get this way an effective density possibly as large as 15,000~kg/m$^3$.  (For this purpose a tungsten heavy alloy will most probably be preferred because it has similar density and yield strength, but a three times larger elastic modulus $E=360$~GPa, than the best uranium alloy \cite{MAGNE1990-,LANZ-2001-}.)  Second, in order to get the desired factor of five increase in effective opacity, one can double the length of the penetrator so that $L=7.8$~m, and the total weight becomes $2,300 \times 5 = 11,500$~kg.

   Will such a design work?  Most probably yes, provided the 8-meter-long, 10-ton-heavy penetrator can be delivered and properly guided to the target, which seems possible since its aspect ratio of $7.8/0.36 \approx 22$ is similar to that of modern kinetic-energy antitank penetrators \cite{LANZ-2001-}.  But what about the deceleration on impact --- will it differ from that of B61-11 for which Eq.~\eqref{rob:1} gave an average value of 6,000~g assuming a penetration $D=2$~m?  As a matter of fact, no: for a penetration of 30~m Eq.~\eqref{rob:2} actually gives a mean deceleration of only 400~$g$.  Indeed, the maximum impact pressure, which can be estimated by Eq.~\eqref{rob:1}, does not depend on the penetration depth or the penetrator's weight, only on its length $L$ so that the mean deceleration is smaller for a greater penetration depth.  Moreover, by going to the penetrator's rest-frame, one sees that the target is always impacting the penetrator with the ram pressure $p = 1/2 \rho_t v_p^2$, which does not depend on any of the penetrator's characteristics except velocity.  Thus, a warhead that survived penetration after free-fall from a given height, will also work for a heavier and longer penetrator able to penetrate deeper into the ground after falling from the same height.

   In summary, by simply increasing the effective opacity of the penetrator one can --- in principle --- achieve any desired penetration, and the existing B61-7 warhead can be used as the explosive charge with only minor modifications.  However, the actual delivery, guidance, and successful penetration of the resulting very-heavy and very-long penetrator will pose considerable engineering problems, which are out of the scope of the present paper.

\subsection{Maximization of impact velocity}
\label{sci.2:0}

As shown by Eqs.~\eqref{fun:12} or \eqref{pen:1}, the second general possibility to increase penetration is to augment the impact velocity.  This can be done for example with a rocket-engine fitted to the bomb, which has the advantage that the weapon could be released by a low-flying aircraft rather than dropped from a stratospheric bomber.\footnote{One could, alternatively, drop the bomb from an orbiting platform: a velocity of some 3 km/s would require an initial altitude of 460 km and a fall time of five minutes.  Dropping long-rod penetrators from outer-space is a commonly discussed space weapon, see, e.g., \cite[p.~70--72]{DEBLO2004-}.}

However, it is not possible to increase the impact velocity beyond the material limit set by the condition that the penetrator has to remain intact upon impact and during penetration, a limit that is defined by the yield (or compressive) strength of the penetrator casing.  Moreover, as will be shown below, increased velocity also means increased maximal impact deceleration, which results in severe constraints on the nuclear warhead and its ancillary components.  For these reasons, a realistic design will most probably be a compromise in which the impact velocity is increased at the same time as the effective opacity of the penetrator.  However, in order to simplify the argument, we will assume in this subsection that the primary free parameter is the impact velocity, and that everything is done to make its increase possible.

Let us rewrite Eq.~\eqref{fun:12} in the form
\begin{equation}\label{sci:1}
    D \propto \rho_p L \Bigl( \frac{v_p(0)}{\sqrt{2\rho_t Y_t}}
                           -  \frac{     1}{           2\rho_t}  \Bigr),
\end{equation}
where the `proportional to' symbol has been introduced in order to stress that what we are interested in is the scaling of the penetration depth with the materials's properties, and where the maximum initial velocity consistent with the compressive strength of the penetrator is approximately given by Eq.~\eqref{fun:14}, i.e.,
\begin{equation}\label{sci:2}
         v_{p,\text{max}}(0) \approx \sqrt{\frac{2Y_p}{\rho_t}}.
\end{equation}

  Let us consider again a concrete target, and suppose that we accept as a starting point a 4~m-long steel penetrator similar to the GBU-28.  Since it is capable of penetrating 6~m of concrete at an impact velocity of 0.5~km/s, we ask for a five time increase in velocity to $v_p(0) = 2.5$~km/s in order to reach $D = 30$~m according to Eq.~\eqref{sci:1}. It then follows from Eq.~\eqref{sci:2} that the compressive strength of the casing should be at least equal to 8~GPa, i.e., about three times larger than the strongest steel alloy available, maraging steel (Mar 350) which has a yield strength of 2.7~GPa.  

   We have therefore to consider a material different from steel for the casing (or at least for its outer skin).  If for example we take a ceramic of the type used in tank armor \cite[Table~2]{LUNDB2004-}, we find that either boron carbide (B$_4$C, $Y_p = 15.8$, $E=460$~GPa, $\rho_p =2490$) or silicon carbide (SiC, $Y_p = 12.1$, $E=450$~GPa, $\rho_p =3230$) could be used --- provided their lower density is compensated for by replacing some of the steel in the penetrator by tungsten or uranium in order to keep the same effective density. 

  Therefore, provided the penetrator could be jacketed with a ceramic or some other superhard and elastic material, it may survive the factor of $5^2 = 25$ increased impact pressure implied by a 5 times larger penetration depth.  But what about the the nuclear warhead and its ancillary components?

  Let us rewrite Eq.~\eqref{rob:2} --- which gives an estimate of the maximum deceleration of a perfectly hard and rigid projectile --- in terms of $v_p^2$ instead of the ballistic range, i.e., 
\begin{equation}\label{sci:3}
      a_{\text{max}} \approx \frac{v_p^2}{2L}.
\end{equation}
With $L=4$~m and $v_p=0.5$~km/s, this is 31~km/s$^2 = 3,200~g$, a deceleration smaller than the 15,000~$g$ that we can assume the B61-7 can survive.  However, at a velocity $v_p=2.5$~km/s, the maximum deceleration becomes 80,000~$g$, i.e, significantly larger than this value.

  Returning to the analysis made in Sec.~\ref{rob:0} in which we concluded that all the major nuclear components in the B61-7 are axially symmetric and strongly held in position, we can infer that at least all its metallic (uranium, plutonium, aluminum, etc.) and homogeneous (beryllium, lithium-deuteride, etc.) components may be able to survive a deceleration of possibly up to 100,000~$g$.  On the other hand, the high-explosives (which have a heterogeneous and fragile constitution) are known to become brittle and to crack at sustained accelerations greater than 30,000~$g$ \cite{LANZER1981-}.  This can cause significant problems such as premature ignition and non-perfectly-symmetrical implosion.

  Moreover, as is well known, the many ancillary components that are necessary for the proper functioning of a nuclear weapon are in fact much more fragile than its main nuclear components.  These include the safety and security devices, the neutron generators, fuzes, firing systems, high-explosive detonators, etc. It is primarily the resistance to failure of these key components that has to be improved to make them compatible with the increased stresses stemming from a twenty-five-fold increase in impact stress.  

   At that point we see that if deeper penetration has to succeed by going to higher impact velocities, the primary implication is that considerable progress has to be made in material science (i.e., to produce new superhard materials suitable for strengthening the weapon's casing, and new high-explosives than can sustain greater accelerations) and in micro-electromechanical engineering (i.e., to build stronger ancillary components).  In both of these engineering domains, the current ``buzz word'' is nanotechnology \cite{GSPON2002B}, which is a term emphasizing that improving materials is mainly to be able to design them by controlling their structure at the atomic or molecular level (better alloys, composite materials, and chemical explosives); and to be able to reduce as much as possible the linear size of all moving components because (at a given yield stress) this is the easiest way to improve resistance to mechanical stress.  Indeed, if using elementary elasticity theory we calculate the bending of rod (fixed at one extremity) of length $\ell$ and density $\rho$ under the effect of an acceleration $a$ one finds the relation
\begin{equation}\label{sci:4}
      \text{stress} = \frac{\text{force}}{\text{surface}}
               \approx \rho \ell a  <  \text{maximum stress} \approx Y_s,
\end{equation}
which implies that smaller objects have a larger resistance to stress than larger ones.

   In summary, by increasing the impact velocity by a factor of five one is reaching a number of material and engineering limits which are at the forefront of contemporary research and development.  If a compromise is made and the earth penetrating weapon is made somewhat heavier and longer than the circa 2,000~kg and 4~m we took as a starting point, the resulting weapon could not be much lighter and smaller than the design considered in the previous subsection, especially if we take into account the weight and volume of the rocket motor required to increase impact velocity.

\subsection{Active penetration by conventional means}
\label{sci.3:0}

After having considered the possibility of higher velocities and ``wrapping the weapon in some superhard new nanomaterial,'' the third general possibility to increase penetration is to use  ``repeated explosions to clear the path'' through the ground \cite[p.33]{MALAK2003-}.

 This possibility is based on the idea that weakening a target before impact could increase penetration depth.   In conventional weaponry this is the rationale of ``tandem'' or ``two-stage'' warheads in which a precursor shaped-charge jet is followed by a kinetic-energy projectile containing a high-explosive filling \cite{HEIDE2001-}.

  To evaluate the influence of a precursor hole, the best is to first assume that the hole has been ``pre-drilled,'' i.e., that some target material has simply been removed before impact of the penetrator, but that otherwise the target material has not been weakened.  In that case the theory developed in Sec.~\ref{fun:0} still applies, provided one corrects the penetration given by Eq.~\eqref{fun:12} in inverse proportion to the ratio of the area of the circle defined by the projectile radius $a$, and that of the annulus of outer radius $a$ and inner radius $b$, the pre-drilled hole's radius \cite{TELAN2001-}, 
\begin{equation}\label{sci:5}
      D(a,b) = \frac{a^2-b^2}{a^2} D(a,b=0).
\end{equation}
This expression has been found to be in reasonable agreement with various experiments \cite{TELAN2001-}, which showed that it generally tends to underestimate the actual penetration depth when $b/a$, the relative radius, is larger than $\approx 0.3$, i.e., when the pre-drilled hole radius is not much smaller than the penetrators's radius.  Thus, as the increased penetration given by Eq.~\eqref{sci:5} depends on the square of the relative radius, a small pre-drilled hole has only a very small influence on penetration depth.

  It is therefore important to evaluate the influence of the target's material weakening when, for instance, the hole is made by means of a precursor shaped-charged and the material is damaged several crater radii away from axis during penetration of the jet.  This can be done by integrated experiments  (provided they can be made), by separated experiments in which the shaped charge and the kinetic penetrator strike the target in different experiments, or by computer simulations calibrated with data from such experiments.  For example, in one such simulation backed by experiment \cite{HEIDE2001-}, it was found that whereas the penetration depth of a caliber 60~mm projectile impacting a concrete target at a velocity of 0.51~km/s is 115~cm, the same projectile striking the target pre-drilled and weakened by a shaped charged charge jet to a depth of 60~cm will exit a 140~cm thick target with a residual velocity of 0.23~km/s.

It is therefore possible to aid kinetic-energy penetration by means of a precursor shaped charge, but the improvement is not dramatic.  In particular, leaving aside the complex interactions (such as slowing-down effects\footnote{These effects can be weakened by spacing the shaped charge and the projectile.}) of the precursor charge on the follow-through projectile, one may estimate that a ``tandem'' penetrator may have twice the penetration capability of a kinetic-energy penetrator on its own.  If now the process is repeated and a hypothetical second shaped charge is detonated underground (once a first shaped charge followed by a first penetrator have opened the way to some depth) one could go further down  --- but at the expense of great complexity and a very uncertain design. 

   Finally, if we consider a single tandem such that a shaped charge would aid the penetration of a weapon containing a nuclear warhead like the B61-7, which has a minimum diameter of about 30~cm, the precursor hole should have a diameter of at least 5 to 10~cm over a length of more than 5~m to have some positive effect.  This implies a very large and powerful shaped charge jet, which is very difficult to create using conventional means, because tens (or possibly a few hundreds) of kilograms of high-explosives may be required to form and drive such a jet.\footnote{In theory, a sufficiently powerful jet could be driven by the X-rays from a low-yield nuclear explosion \cite{GSPON2005-}. But the resulting third-generation ``DEW plus EPW'' tandem-warhead appears to very difficult to develop without extensive nuclear testing, and would imply a low-altitude nuclear detonation before penetration.}

\subsection{Active penetration by thermonuclear means}
\label{sci.4:0}

The fourth general possibility to increase penetration is to look at the contemporary trend in nuclear weapons development, and to see whether some radically new technology could be on the design horizon that would complete change the rules of the game.

As a matter of fact, since many years and more so since the conclusion of the Comprehensive Test Ban Treaty, nuclear weapons's testing has moved into laboratories where huge laser facilities are used to better understand the ignition and burn of small thermonuclear explosions.  The two largest such facilities are under construction in the United States (the National Ignition facility, NIF) and in France (the Laser M\'egajoule, LMJ).  But quite powerful facilities are already or will be soon available in Japan, England, China, Russia, Germany, etc.

The full understanding of the ignition and burn of small pellets of fusion material (the so-call Inertial Confinement Fusion, ICF,\footnote{For a recent review of ICF physics see Ref.~\cite{LINDL2004-}, and for a recent appreciation of the progress on NIF see Ref.~\cite{MOSES2005-}.} route to thermonuclear energy) will open the way to a radically new generation of nuclear weapons in which there will be no, or very little, fissile material \cite{GSPON1997-}.  These ``fourth-generation nuclear weapons'' (FGNW) are a great concern for nuclear proliferation because they can potentially be built by all technologically advanced countries, including non-nuclear-weapon States, and because they are militarily ``very sweet'' as they are very powerful and politically usable --- since they do not have some of the major drawbacks of existing types of nuclear weapons: radioactive fallout and too much yield.

For the purpose of the present paper we will assume that such fourth-generation nuclear explosives can be built, and that such explosives with yields in the range of 1~to 1000 \emph{tons} of high-explosive equivalent could become available at most a few decades from now.  We will also simplify the discussion by assuming that these explosives can be approximated by a point source of high-energy (i.e., 14~MeV) neutrons, even though they will most probably consist of spherical or cylindrical devices with a radius of at least a few centimeters.  

   The availability of such powerful and compact sources of neutrons leads to many military applications other than just explosives.  This is because the neutrons can be used to work on some material which can be accelerated (as by a rocket engine) or forged and projected (as by a shaped charge) \cite{GSPON2005-}.  However, because of their small size (say a diameter of 5~cm or less instead of the 15~cm of the smallest nuclear artillery shells), fourth-generation nuclear warheads could already take more easily advantage of a precursor chemical-explosive shaped-charge in penetrating and defeating a bunker or buried target than current nuclear warheads.

In the context of the present problem of driving a powerful explosive deep into the ground before detonating it, FGNW technology gives the possibility to envisage active earth-penetrating devices that could be both compact and lightweight.  Although the technological and engineering hurdles could be formidable, it is essential not to be shy because the consequences can be far-reaching, and further nuclear weapons proliferation is at stakes.

For example, one can think of a tandem weapon in which a front pure-neutron explosive of 10~\emph{tons} would create a powerful shaped charge jet, and a back pure-neutron explosive of 100~\emph{tons} would drive a package containing the main nuclear explosive charge into the hole formed by the jet.  The development of such a complex device would require very powerful computer simulation capabilities, and considerable interaction with reduced scale experiments at ICF facilities.\footnote{A fundamental difference between the fission and the fusion explosive processes is that fusion is scaleable while fission is not: fission is bound to high-yield explosions by the critical mass.}  In fact, the development of FGNWs such as the earth-penetrating device just described might be a far more credible justification than ``stockpile stewardship'' for the enormous computer facilities made available to the nuclear weapons laboratories...

In summary, new types of nuclear explosives in which the main output of energy is in the form high-energy neutrons may lead to entirely new solutions to the deep earth penetration problem.  However, for these solutions to become a reality, considerable research and development is required to improve the non-nuclear components of the penetrators as well, just like for the more conventional approaches to increased penetration, as was seen in the previous subsections.  This implies that the RNEP program is consistent with the development of FGNWs and their use in advanced earth penetrating weapons.

\section{Environmental and political limitations}
\label{env:0}

The environmental consequencies of shallow buried nuclear explosions have been addressed in a number of recent studies, and it has been convincingly argued that insufficiently deep burial of an earth penetrating warhead will lead to extensive radioactive pollution of the environment \cite{NELSO2002-, NAS--2005-}.  

   For example, the minimum depth of burial required to fully contain a 1~\emph{kt} nuclear explosion at the Nevada Test Site is about 100~m. The implication for the case under consideration in this paper, i.e., a penetration depth of about 30~m in concrete or rock, or about 100 to 150~m in sand or earth, is that the nuclear yield should be less than 1~\emph{kt} if radioactive fallout is to be avoided.

Therefore, if a buried target can only be defeated by an explosive yield on the order of the maximum that can delivered by the B61-7, i.e., 340~\emph{kt}, one cannot expect that this explosion will be contained.  The use of a RNEP will thus have the same environmental and political consequences as any atmospheric nuclear explosion, and will not in any way be comparable to a kind of ``nuclear test explosion on enemy territory in which there would be no visible mushroom cloud.''

    Another seldom mentioned political limitation to the use of earth penetrating nuclear weapons arises even if the explosions are fully contained.  This is because any nuclear weapon detonated below the surface of the Earth creates a radioactive hot spot, containing the highly-radioactive fission-products, as well as the fissile materials that have not been fissioned during the explosion --- that is more than 90\% of the initial fissile-materials content of the warhead.\footnote{In the case of an air-burst, the fission products and non-fissioned fissile materials are dispersed over a large surface or carried away in the form of nuclear fallout.}  This leads to numerous problems, especially if the weapon contained plutonium.  Indeed, most of the un-fissioned plutonium left in the cavity created by an underground nuclear explosion can later be recovered, possibly by terrorists, and reused to manufacture nuclear weapons.\footnote{The possibility of recovering plutonium left-over (or intentionally bred --- as well as other actinides or isotopes) in underground nuclear explosions, has been studied during the 1960s in the framework of the ``Plowshare'' program, e.g., \cite{KARRA1965-}.  Interest in the concept has persisted in various forms until the present, e.g., \cite{SAHIN2003-}.} This is the problem of \emph{``plutonium mines,''} which are a major nuclear proliferation concern in countries such as Kazakhstan \cite{STONE2003-} and Algeria, where numerous plutonium-based nuclear explosives were detonated underground.  

Consequently, earth penetrating warheads like the B61-7 use highly-enriched uranium as fissile material, because in that case the un-fissioned U-235 cannot be recovered after the explosion as it will be intimately mixed with the U-238 from the non-enriched uranium components of the weapon.\footnote{According to Ref.~\cite{SUBLE1997-} the B61 primary consists of beryllium reflected plutonium. If that is the case, any site of use of a B61-11 or future RNEP will have to be cleaned up, or else monitored for thousands of years like all other ``plutonium mines.''} The only plutonium will be the small amount bred by neutron capture in U-238.  On the other hand, the un-burnt fraction of the tritium used for boosting is much less a problem.  First because tritium is very volatile and easily drifts away from the point of explosion, second because its half-life is only twelve years, and third because its radio-toxicity is relatively small.  For these reasons, in the case of FGNWs, the related problem of ``tritium mines'' is potentially much less serious than the problem of ``plutonium mines.''

\section{Discussion}
\label{dis:0}

The paper was organized to begin with a derivation, from first principles, of the fundamental equation giving to first order the penetration depth of a rigid penetrator into a solid material.  This derivation was necessary in order to confirm that in the parameter regime of interest to defeat well-protected underground targets there is very little room for improvement outside the main parameters which appear explicitly in Eq.~\eqref{fun:12}.  In particular, fine tuning parameters such as the shape of the penetrator, or its surface texture, will only have second order effects on penetration depth, at the level of 10\% or less. 

We have then reviewed the physical and engineering reasons which explain the  penetration depth of the main earth-penetrating nuclear weapon currently available in the United States's arsenal, the B61-11, and derived crude estimates in terms of maximum sustainable decelerations of the ultimate impact-resistance of the B61-7, the warhead enclosed in the B61-11 penetrator.

This has led us to consider four different possibilities for designing a RNEP able to bury the B61-7 warhead 30 meters into concrete, or 100 to 150 meters into earth.  The first solution is simply to build a 8-meter-long, 360-mm-caliber, 10-ton-heavy penetrator made out of a good tungsten alloy.  This would be a ``1950s generation'' style weapon, albeit fitted with a modern two-stage thermonuclear warhead, and a GPS guidance system allowing for high precision delivery under all weather conditions.  Little sophisticated engineering development would be required to produced the penetrator, and no change would have to be made in the B61-7 warhead: the version carried by the B61-11 could be used without modification.

However, considering the weight and size of that first solution, we have investigated the possibility of augmenting the impact velocity by means of a rocket engine.  This immediately led to contemplate considerable engineering difficulties, because while the major nuclear components could probably resist the increased stresses arising from the greater impact and penetration decelerations (with the exception of the chemical explosives which may have to be improved), many of the more fragile ancillary non-nuclear components of the weapon would have to be redesigned, using new materials and techniques which are not yet readily available.  Moreover, even by making a compromise such that the penetrator would not be so large and heavy as the one considered in the first possibility, fitting it with a suitable rocket engine would not make it much less cumbersome.

We have therefore turned to the possibilities of aiding penetration by means of a precursor shaped-charge.  Here we have met again with considerable engineering difficulties, especially since designing a shaped charge powerful enough to clear the way to a circa 30~cm diameter penetrator containing a B61-7 warhead looks very difficult, and this only to augment its penetration capability by possibly 50\% under the best conditions. 

The last possibility we have considered is more speculative:  Assuming that radically new types of compact nuclear explosives will become available in the next few decades (i.e., low-yield pure-fusion thermonuclear explosives derived from ICF pellets which are extensively studied at present in many laboratories) speculations have been made on hypothetical new designs which could provide a technically more attractive solution than the three previous possibilities.  In particular we have taken the example of an earth-penetrating weapon which would be equipped with a powerful thermonuclear-driven precursor shaped-charge, followed by a main charge that would be accelerated and driven into the ground by the momentum supplied by a 100 \emph{tons} thermonuclear explosion at the back of it.

The main conclusion stemming from considering this more speculative possibility is that the ancillary technological advances required in order to make it feasible are similar to those implied by the second possibility, i.e., considerable advances in the realms of materials science,  micro-electromechanical systems's engineering, and nanotechnology.  Since these potential advances significantly overlap with the requirements of many other future weapons, both conventional and nuclear, it is therefore possible that an important reason for considering an improved version of the B61-11 is simply to develop these technologies as they will be needed, for instance, to weaponize fourth generation nuclear explosives.

Finally, we have briefly looked at some major environmental and political implications of the actual use of earth-penetrating nuclear weapons.  Since these aspects have been discussed at length in many recent publications --- therefore reviving the still open debate on the arguable military utility and political roles of nuclear weapons --- we have not gone further than mentioning that even if an underground nuclear explosion is fully contained, the site of the explosion will in due time have to be clean-up, in order to avoid long-term contamination and further nuclear proliferation problems if someone tries to recover the un-burnt fissile materials remaining after the bomb's explosion.\footnote{There are also other problems with the used of RNEPs, such as the probabilistic nature of projectile penetration \cite{SIDDI2001-} and assessing the actual damage made to the intended target \cite{VALEN2005-}.}

\section*{Acknowledgments}

The author is indebted to David Hambling for stimulating correspondence, to John Coster-Mullen for several contributions to the iconography, and to Dr.\ Jean-Pierre Hurni and many others for their help, ideas, and criticism.

\section{Appendix: The ``Young/Sandia penetration equations''}
\label{app:0}

A reviewer has asked why there was no explicit reference in this paper to the ``Young/Sandia penetration equations'' which are widely used in the U.S.\ weapons community, i.e., Ref.~\cite{YOUNG1967-, YOUNG1997-} and references therein.  The main reason for this was that these equations are in fact  empirical engineering formulas providing a good estimate for target penetrability in materials for which measurements have been made using projectiles that are typical of those in existence in U.S.\ arsenals ---  that is, regularly updated formulas providing a simple and good fit to an expanding data base. For these reasons the ``Young/Sandia penetration equations'' cannot reliably be used outside of the range of materials, projectiles, and parameters corresponding to this data base.

  Nevertheless it is useful to compare the approximate but physically well understood formula derived in Sec.~\ref{fun:0}, i.e., Eq.~\eqref{pen:1} in the case of a typical concrete target, to the main ``Young/Sandia penetration equation,'' i.e., Eq.~(A-3.2) on page A-2 of \cite{YOUNG1997-}.  This equation can be written
\begin{equation}\label{app:1}
      D \approx  9.63 \times S \times N \times \Bigl( L \frac{\rho_{\text{eff}}}
             {\rho_{\text{Fe} }} \Bigr)^{0.7} (v_p\text{[km/s]} - 0.0305),
\end{equation}
where for the sake of comparison the velocity is measured in km/s as in Eq.~\eqref{pen:1}.

  In this formula, the exponent of the projectile's opacity (``weight to area ratio'' in engineering language) has evolved for 0.5 in the 1967 version to 1.0 in the 1988 version, see Ref.~\cite[p.~3]{YOUNG1997-} and Eq.~(2) in Ref.~\cite{NELSO2002-}, to finally become 0.7 in the current version, i.e., Eq.~\eqref{app:1}.  However, assuming for example a ``nose performance factor'' of $N=0.72$ for a moderately sharp nose, and a ``S-factor'' of $N=0.72$ for a relatively hard concrete, we get $ 9.63 \times S \times N \approx 5$ as in our Eq.~\eqref{pen:1}.  Therefore, in that case, for a projectile of length on the order of a fraction or a few meters, Eq.~\eqref{pen:1} and Eq.~\eqref{app:1} will given comparable estimates for penetration depth.

It can therefore be concluded that Eqs.~\eqref{pen:1} and \eqref{app:1} will give comparable results for typical standard projectiles and target materials, but that the ``Young/Sandia penetration equation'' \eqref{app:1} will typically be more accurate in these cases.  However, if one considers much larger velocities or projectile lengths, or else very untypical materials, Eq.~\eqref{pen:1} may give more reliable estimates, especially if the goal is to compare relative penetration performances rather than absolute penetration depth estimates.

Finally, reference \cite{YOUNG1997-} and the ``S-factor'' are quite useful in providing first order estimates for the typical relative penetration depths into various natural materials such as rock, sand, clay, soil, or ice.  In particular, if a moderately soft concrete can be characterized by $S \approx 1$, ice and frozen soil may be  characterized by $S \approx 4$, hard soil by $S\approx 5$, and soil fill (as above a buried bunker) by $S\approx 8$, etc.

\end{document}